\documentstyle[12pt]{article}
\def\IJMP #1 #2 #3 {{\it Int.\ J.\ Mod.\ Phys.}\ {\bf #1}\ (#2) #3}
\def\MPL #1 #2 #3 {{\it Mod.\ Phys.\ Lett.}\ {\bf #1}\ (#2) #3}
\def\NC #1 #2 #3 {{\it Nuovo Cim.}\ {\bf #1} (#2) #3}
\def\NP #1 #2 #3 {{\it Nucl.\ Phys.}\ {\bf #1}\ (#2) #3}
\def\PL #1 #2 #3 {{\it Phys.\ Lett.}\ {\bf #1}\ (#2) #3}
\def\PR #1 #2 #3 {{\it Phys.\ Rev.}\ {\bf #1}\ (#2) #3}
\def\PP #1 #2 #3 {{\it Phys.\ Rep.}\ {\bf #1}\ (#2) #3}
\def\PRL #1 #2 #3 {{\it Phys.\ Rev.\ Lett.}\ {\bf #1}\ (#2) #3}
\def\RMP #1 #2 #3 {{\it Rev.\ Mod.\ Phys.}\ {\bf #1}\ (#2) #3}
\def\ZP #1 #2 #3 {{\it Z.\ Phys.}\ {\bf #1}\ (#2) #3}

\def\ps{J/\psi}
\begin{document}
\begin{flushright}
hep-ph/9608296
\end{flushright}
\vspace*{1cm}

\begin{center}
{\bf Color-Octet Contribution to $\ps$ Hadroproduction with Nonzero $p_T$
at Fixed Target Energies}\\
\vspace*{5mm}
{L.~Slepchenko, A.~Tkabladze}\\
{\it Bogoliubov Laboratory of Theoretical Physics,}\\
{\it JINR, Dubna, Moscow Region, 141980, Russia}\\
\end{center}

\begin{abstract}
We calculated the color-octet contribution to the $\ps$ hadroproduction
at fixed target energies $\sqrt(s)\simeq40$ GeV. We consider the $\ps$
production with transverse momenta which can not be explained by
primordial motion of partons, $p_T>1.5$ GeV. It is shown that
color octet contribution is dominant at these  energies and reduces
large dicrepancies between experimental data and color singlet model
predictions. Having taken into account both contribution one needs a K-factor
 about 2-3  to explain the experimental data.
\end{abstract}

\newpage

Investigation of the processes of heavy quorkonia production gives an
excellent possibility  to study a perturbative feature of QCD and allows one
to extract long-distanse  effects connected with hadronization phaze.
Due to the large mass of $c$ and $b$ quarks production of the heavy
 quark-antiquark  pair takes place at the short distances ($\sim m_Q$) and
can be controled within the framework of perturbative QCD (PQCD).
Then, the quark pair is bound into a quarkonium in a time scale of
order of the the inverse binding energy, $\tau\simeq1/(m_Qv^2)$, where
v is the velocity of quarks within the bound state. 
As $m_Q\to \infty$, the heavy quark velocity inside the bound state
$v\sim 1/ln(m_q)\to 0$, and above mentioned two time scales become widely
separated $m_Q>m_Q v^2$ (for $b\bar b$-system $v^2\simeq0.08$ and for
$c\bar c$ -- $v^2\simeq0.23-0.3$, charmonium is not truly nonrelativistic
system). The separation of time scales makes possible to factorize
production process of a heavy quarkonium in the short distance and long
distance parts.
 Short distance part describes the heavy quark-antiquark pair
production  and can be controled perturbatively. 

 The long distance part
is related   to
matrix element responsible for the hadron formation from the quark-antiquark
pair.

 Quarkonium production has traditionally been calculated in the color
singlet model (CSM) \cite{CSM}. In this approach is proposed that
 quark-antiquark 
pair  is  produced in a color singlet state with the quantum numbers
of the corresponding hadron.
 This heavy pairs eventually creates the hadronic
state with a probability determined by the appropriate quarkonium wave
function  at the origin. It is assumed that for heavy
quarks soft gluon emmision as well as other nonperturbative effects such
 as higher
twist contributions are negligible.  While this model gives a reasonable
description of $\ps$ production cross section  shape over
 $p_T$ or $x_F$ it completely fails in the explanation of the integrated cross
cross section (K factor 7-10 is needed to explain experimental data).
The anomalously large cross section \cite{CDF} of $\ps$ production at large
 transverse
momenta at the Tevatron reveals another bad feature of the CSM.
Within the framework of the CSM it is impossible to explain the anomalously
large $\psi$ \cite{7GS} and direct $\ps$ production \cite{CAC}
at  CDF experiment.

Indeed, the requirements of the CSM is very strong and lead to the
suppression of both fusion and fragmentation contributions. Production
of a  quark-antiquark pair on the time scale $\tau\simeq1/m_Q$ with the proper
quantum numbers of subsequent hadronic state must be accompanied with the
emmition of hard gluons. Therefore, production  cross section in the CSM
is suppressed by powers of $\alpha_s/\pi$.

The CSM is a  nonrelativistic model where the relative velocity between
the heavy constituents in the bound state is neglected. But discrepancies
between experimental data and CSM predictions hint that $O(v)$ corrections as
well as other mechanisms of quarkonia production, which do not appear at
leading order in $v$, should be considered. 
Treating higher order in
$v$ it is possible to avoid the supression caused by emission of hard gluons,
 if one allows a quark-antiquark pair to be produced on the time scale
  $1/m_Q$ in  any color state (singlet or octet) with any quantum numbers.
Such a pair evolves into a hadron state by emitting of soft gluons
 with momenta of an order of $m_Q v^2$.
 Expansion of quarkonium cross sections and decay widths
in the powers of relative velocity $v$ of  heavy quarks in a bound
state has recently been realized
 in terms of Nonrelativistic QCD (NRQCD) \cite{NRQCD}.
This formalism implies not only color-singlet processes but so called
color-octet mechanism, when a quark-antiquark pair is produced on the small
time scales in the color octet state and evolves into hadron by emmition
of soft gluons. According to the factorization approach based on the NRQCD,
the production cross section for quarkonium state  H in the process
\begin{eqnarray}
 A+B\to H+X
\end{eqnarray}
can be written as
\begin{eqnarray}
  \sigma_{ij} = \sum_{i,j}{\int_{0}^{1} {dx_1 dx_2 f_{i/A}(x_1) f_{j/B}(x_2)
\hat\sigma(ij\to H)}}\\
\hat\sigma(ij\to H) = \sum_{n}{C^{ij}[n]<0|O^{H}[n]|0>}\nonumber
\end{eqnarray}
where the $f_{i/A}$ is the distribution function of the 
parton i in the hadron A.
Subprosess  cross section is separeted into two: short distance, $C^{ij}[n]$,
and long distance, $<0|O^H[n]|0>$, parts. The $C^{ij}[n]$ is the production
cross section of a heavy quark-antiquark pair in the i and j parton fusion.
It should be calculated in the framework of PQCD. The [n] state can be
either a color
singlet or octet state. The $<0|O^H[n]|0>$ describes evolution
of a quark-antiquark pair into a hadronic state. These matrix elements cannot
be computed perturbatively. 
But relative importance of long distance matrix
elements in powers of velocity $v$ can be estimated by using
the NRQCD velocity scaling rules \cite{7KRA}.

The shapes of the $p_T$ distribution of short distance  matrix elements
within the color octet model 
indicates that new mechanism can explain the Tevatron data of direct
$\ps$ and $\psi$ production at large $p_T$.
But unlike color-singlet matrix elements connected to the subsequent
hadronic nonrelativistic wave functions at the origin, the color octet long
distance matrix elements are unknown and should be extracted from 
experimental data. Explanation of the CDF data of S state charmonia
production at the large $p_T$   will be succesfull after testing the values of
color octet long distance matrix elements in  other experimental data

The color octet contribution to the $\ps$ photoproduction was analized
in the papers \cite{KRA,FLE}. Recently, the $\ps$ hadroproduction at fixed
target energies has been studied by including color-octet mechanism
\cite{GS,BR}.  Large discrepancies between experimental data and CSM
predictions for  the $\ps$ production total cross section were explained.
 The color octet contribution is dominant in the $\ps$ hadroproduction at
energies $\sqrt(s)\simeq30-60$ GeV.
The analises carried out in these papers \cite{KRA,FLE,GS,BR}
demonstrate that to fit the photoproduction and hadroproduction data at the
low energies require rather small values for color octet matrix
elements than those extracted from a
prompt $\ps$ production at CDF \cite{CL}.

In the present letter we consider the color octet contribution to the $\ps$
hadroproduction with nonzero $p_T$ at fixed target energies. We called
nonzero $p_T$ such transverse momenta which cannot be explained by
primordial motion of partons inside the colliding hadrons. We consider $p_T>1.5$
GeV to calculate the short distance matrix elements perturbatively and avoid
the infrared and collinear divergencies in the production of $^1S_0$ and 
$^3P_{0,2}$ quark-antiquark states.

In the papers \cite{GS,BR} for calculation of the total cross section of
the $\ps$ production only the  subprocesses $2\to1$ were taken into account.
These subprocesses are the lowest order
 in perturbative series over $\alpha_s$
and give the main contribution to the integrated cross section. The
transverse  momenta of produced particles due to internal motion of
partons are of an
order $\Lambda_{QCD}$, and $p_T$ behaviour of the cross section cannot be
controlled perturbativle.
  The subprosses $2\to2$ are the lowest order which contribute to the
$\ps$ production at $p_T>>\Lambda_QCD$.
\begin{eqnarray}
gg\to(c\bar c)g, \nonumber \\
gq\to(c\bar c)q, \\
qq\to(c\bar c)g, \nonumber
\end{eqnarray}
where $(c\bar c)$ denotes any state of heavy quark-antiquark pair.
These subprocesses give small contribution to the total cross section
because additional $\alpha_s$ and the steeply falling down $p_T$ behaviour.
 The cross section of the $\ps$ production can be written as
\begin{eqnarray}
\sigma_{\ps} = \sigma(\ps)_{dir} +
\sum_{J=0,1,2}{Br(\chi_{cJ}\to \ps X)\sigma_{\chi_{cJ}}}+Br(\psi'\to \ps X)
\sigma_{\psi'},
\end{eqnarray}
where in the production of each  quarkonium state is contributed by  both color
singlet and octet states,
\begin{eqnarray}
\sigma(\ps)_{dir} = \sigma_{\ps}^0+\sigma^8_{\ps}+
\sum{\sigma(Q\bar Q(^{2s+1}P_J^{(8)})<0|O_8^{\ps}(^{2s+1}L_J|0>}
\end{eqnarray}
where the sum stands over the states $^3P_{0,1,2}^8 $, $^1S_0^8$ and $^3S_1^8$.
The expressions for differential cross sections
 for color singlet states we take from
\cite{CSM,WG}. For color octet states the short distance matrix elements have
recently been calculated by Cho and Leibovich \cite{CL}.

As concerned  the long distance matrix elements their number should be
reduced by using NRQCD spin symmetry relations:
\begin{eqnarray}
 <0|O_8^H(^3P_J)|0> = (2J+1)<0|O_8^H(^3P_0)|0>,\\
 <0|O_8^{\chi_{cJ}}(^3S_1)|0> = (2J+1)<0|O_8^{\chi_{c0}}(^3S_1)|0>.
\end{eqnarray}

After that  from the parameters which give main contribution
in the cross section only four independent matrix elements remain;
$<0|O_8^{\ps}(^3S_1)|0>$, $<0|O_8^{\chi_{c1}}(^3S_1)|0>$,
$<0|O_8^{\ps}(^3P_0)|0>$ and $<0|O_8^{\ps}(^1S_0)|0>$.
Unfortunatly, values for these matrix elements obtained from  fitting
various experimental data are  different.
There are two different values for the matrix elements
$<0|O_8^{\ps}(^3S_1)|0>$, $<0|O_8^{\chi_{c1}}(^3S_1)|0>$,
extracted from CDF data by two group of authors.
Using only the dominant fragmentation contributions to the $\ps$
 production Caciari et al.\cite{CAC} obtained the following values:
\begin{eqnarray}
<0|O_8^{\ps}(^3S_1)|0>=15\cdot10^{-3} GeV^3, \nonumber\\
<0|O_8^{\chi_{c1}}(^3S_1)|0> = 2.4\cdot10^{-2} GeV^3. \nonumber
\end{eqnarray}
Using the full perturbative expressions for the short distance matrix
elements Cho and Leibovich obtained  smaller values by factor two for 
the above parameters, 
 $6.6\cdot10^{-3} GeV^3$ and $9.8\cdot10^{-3} GeV^3$, respectively.

As for  the other two  parameters  it is possible to
extract only  their combinations from experimental data.
  From charmonium production at
large transverse momenta at CDF Cho  and Leibovich extracted \cite{CL}
\begin{eqnarray}
<0|O_8^{\ps}(^1S_0)|0>+\frac{3}{m_c^2}<0|O_8^{\ps}(^3P_0)|0> =
6.6\cdot10^{-2} GeV^3.
\label{high}
\end{eqnarray}
Different combination of these parameters were extracted from $\ps$
photoproduction data and hadroprodaction data at fixed target energies:
\begin{eqnarray}
<0|O_8^{\ps}(^1S_0)|0>+\frac{7}{m_c^2}<0|O_8^{\ps}(^3P_0)|0> =
2\cdot10^{-2} GeV^3 \cite{BR},\nonumber\\
<0|O_8^{\ps}(^1S_0)|0>+\frac{7}{m_c^2}<0|O_8^{\ps}(^3P_0)|0> =
3\cdot10^{-2} GeV^3 \cite{KRA,FLE}.
\label{low}
\end{eqnarray}
As one can see from   equations  (8) and  (9) 
the photoproduction and hadroproduction
data at low energies are consistent with each  other within the model error.
But there is the large discrepancy between the low, (9),
  and high energy, (8),
values of the parameters.
Solving the system of equations (\ref{high}) and (\ref{low}),
 one will obtain a
negative value for the $<O_8^{\ps}(^3P_0)>$ matrix element.
The reason for such discrepancy may be an overestimation of the value obtained in
the paper \cite{CL} (equation \ref{high}).
In the prompt $\ps$ production at the Tevatron the  $^1S_0^{(8)}$  and
$^3P_J^{(8)}$ states give a dominant contribution at transverse momenta
near $p_T=5$ GeV. The value $6.6\cdot10^{-2} GeV^3$ was extracted just
from this region of $p_T$. 
But at these values of transverse momenta $\ps$ is produced mainly
at small partonic $x$ ($x\sim 10^{-2}\div10^{-3}$) in the gluon-gluon fusion.
Cho and Leibovich in paper \cite{CL} use the partonic parametrization MRSD0
\cite{MRS}.
 But in the above mentioned region of  $x$  the MRSD0 parametrization
 gives rather smaller value for the gluon
distribution function  
 than more realistic parametrization (GRV LO or GRV HO \cite{GRV}).
Using GRV LO parametrization gives 1.55 times large values for the
production cross sections of  $^1S_0$ and $^3P_J$ states. 
So, more realistic value for the  combination (\ref{high}) is
\begin{eqnarray}
<0|O_8^{\ps}(^1S_0)|0>+\frac{3}{m_c^2}<0|O_8^{\ps}(^3P_0)|0> =
4\div4.4\cdot10^{-2} GeV^3.
\label{highnew}
\end{eqnarray}
Consequently, value of $1.4\div1.5$ times large is needed
for the  $<O^{J/\psi}_8(^3S_1)>$ matrix element to explain the
prompt $J/\psi$ production cross section at $p_T\simeq10$ GeV \cite{CL},
\begin{eqnarray}
  <O^{J/\psi}_8(^3S_1)> = 9\div10\cdot10^{-3} GeV^3.
\end{eqnarray}
After these changes the dicrepancy between the 
combinations (9) and (10) would not be large for the radical choice
 $<O_8^{J/\psi}(^3P_0)>=0$.

We presented here, fig.1, results for the differential cross section 
for the set of parameters
taken from \cite{CL}. We asumed that
  $<O_8^{J/\psi}(^1S_0)>=<O_8^{J/\psi}(^3P_0)>$.

In the fig.2 presented the differential cross section for 
another set of parameters obtained after taking into account 
above mentioned corrections:
\begin{eqnarray}
 <O^{J/\psi}_8(^3S_1)> & = &10.5\cdot10^{-3} GeV^3,\nonumber\\
<0|O_8^{\chi_{c1}}(^3S_1)|0> &  = & 9.8\cdot10^{-3} GeV^3, \nonumber\\
 <O_8^{J/\psi}(^1S_0)> & = &3.7\cdot10^{-2} GeV^3, \nonumber\\
 <O_8^{J/\psi}(^3P_0)> & = &0. \nonumber \\
\end{eqnarray}
For  $<O_8^{J/\psi}(^1S_0)>$ we choose the medium value of the
combinations (\ref{low})  and (\ref{highnew}).
For parameters which describe the transition color octet state into a
$\psi'$-meson we use the values from \cite{CL}.
Theoretical predictions are
compared with the experimental data of experiment E689 at FNAL \cite{E689}.
We use the functional form of for $p_T$ distribution from \cite{E689}
to reproduce the curve of experimental data (fig.1,2).
As one can see from figures color octet contribution is dominant. But the
total theoretical prediction are small
 and K-factor about $2\div2.5$ is needed to explain the experimental data.
It is wort to mention  that in our calculations there is
uncertainty  that lead to the
decreasing of the theoretical prediction in the whole region of considered $p_t$.
The relative velocity of $c$ and $\bar c$ quarks in the charmonium is
about $v^2\simeq0.23\div0.3$. This means that soft gluons emmited from
color octet quark-antiquark state while evolution into $\ps$ has an
momentum about $0.7\div1$ GeV ($2 m_c v^2$).  So, it is necessary to
produce color octet
state of quark-antoquark pair  with mass large then $2m_c$. 
In the case of the photoproduction or the total cross section 
in hadroproduction 
at fixed target energies the kimematical effect from the difference between the
mass of $\ps$ and color octet quark-antiquark pair is very large \cite{BR}
since the gluon distribution rises steeply at small x. This reduces
cross sections twice and 'true' matrix elements would therefore be large than 
those extracted by using the small mass of quark-antiquark pair \cite{BR}.
In our case the influence of such effect is not  so large and leads
to the decreasing the cross section by about $25\%$ at $p_t\sim1.5$ GeV
and by about $10\%$ at $p_t\sim3$ GeV. Another uncertainty comes from 
decay process of quark-antiquark pair into $\ps$. 
The $\ps$ production cross section shape over $p_T$ differs from that of
color-octet quark-antiquark pair.
For qualitative estimations of this corrections we assumed that produced
quark-antiquark state  is unpolarized and its decay into $\ps$ and
gluon is isotropic. In this approach we  calculated the deviation
of the $\ps$ $p_T$ distribution shape from the distribution shape of
 heavy quark-antiquark pairs with mass 4 GeV.
The difference is negligible near  $p_T\sim1.5$ GeV and at 3 GeV is about
$20\%$. Unfortunately,
 we can only qualitatively estimate  these two corrections because
the true mass of a quark-antiquark state before evolution into hadronic
state is unknown. But it is obvious that these two corrections lead to
the reduction not to the rising of the cross section.

In conclusion, we have calculated the color octet contribution to the
$\ps$ hadro\-pro\-duction with nonzero $p_T$ at fixed target energies.  Color
octet processes are dominant in the $\ps$ production and large about an order
of magnitude then color singlet contribution.

After taking into account color octet contribution the K-factor
 about $2\div3$ is need to explain the experimental data.
Uncontrolled corrections coming from the decay process of quar-antiquark
states into $\ps$ and 'soft' gluons leads only to the decreasing of
the cross section and discrepancies become  larger.

We are indebted  to N.~Kochelev, W.-D.~Nowak and O.~Teryaev for useful
discussions and helpful comments.

\newpage

\begin{center}
{\bf  FIGURE CAPTIONS}
\end{center}
\noindent
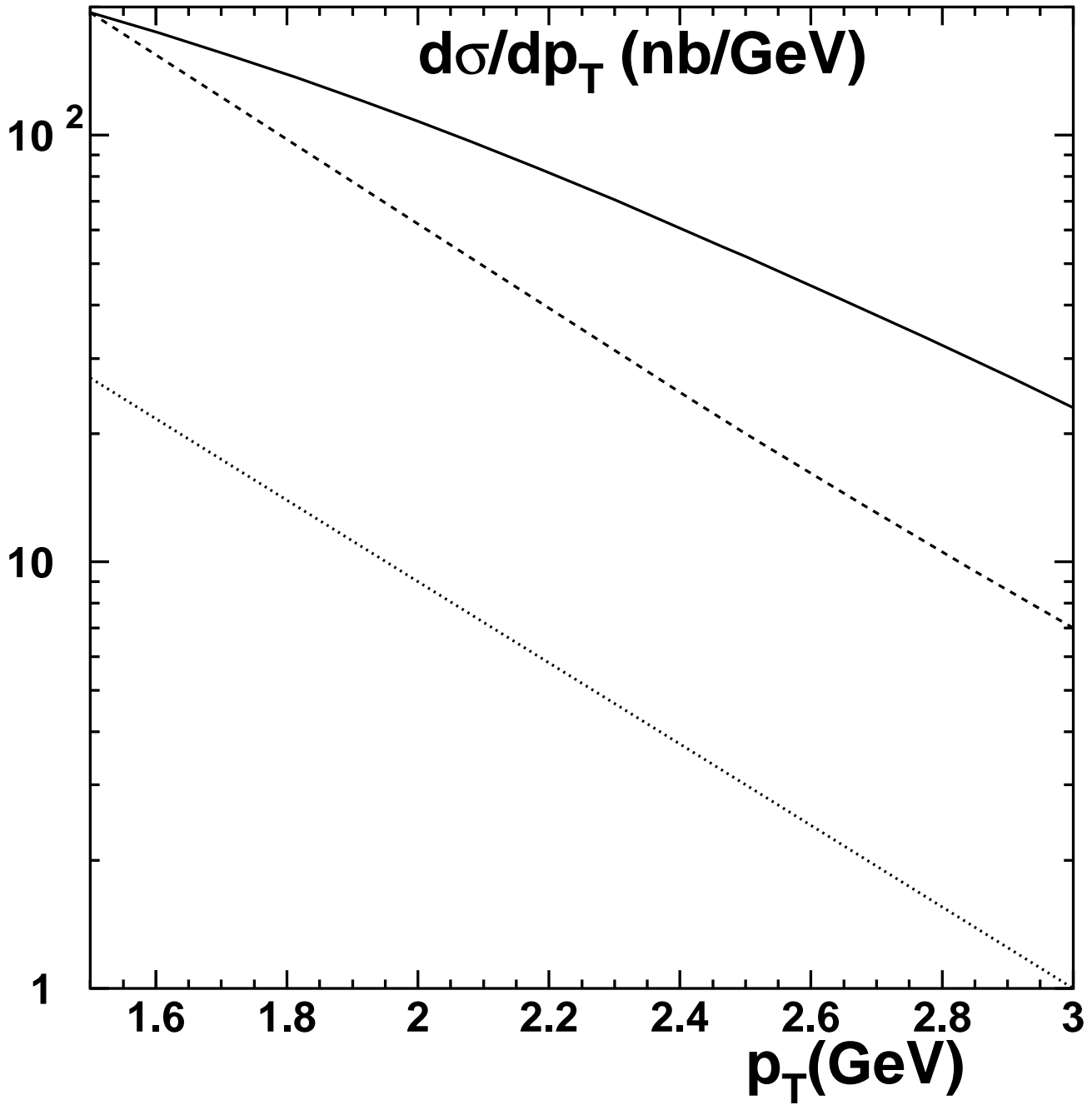
Fig.1. Transverse momentum diferential cross sections for the set of color octet
matrix elements from \cite{CL};
solid curve represents experimental data from \cite{E689}, dashed
curve -- total theoretical predictions and dotted curve -- CSM
predictions.
\vspace{1cm}

\noindent
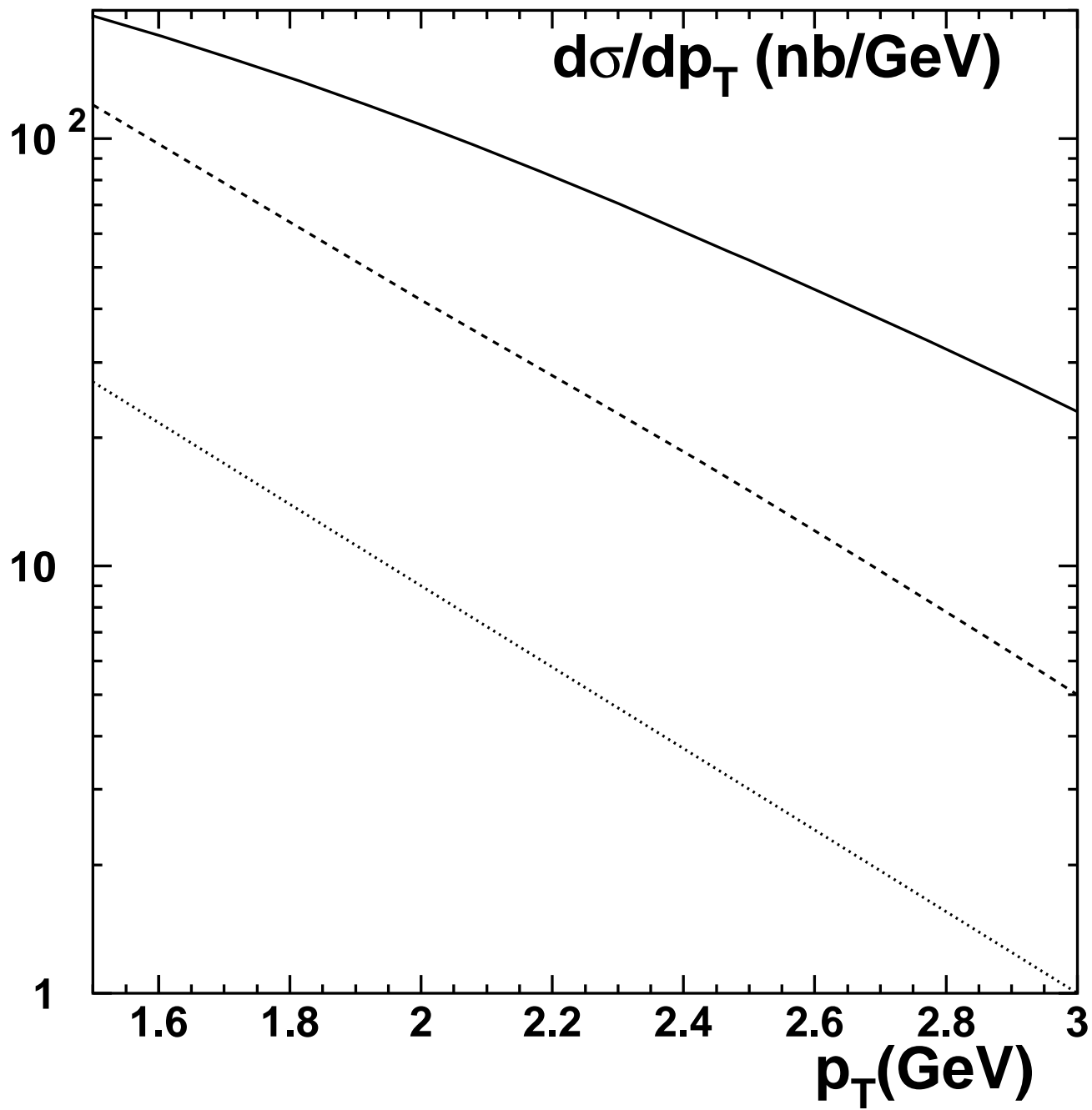
Fig.2. Transverse momentum diferential cross sections for the color octet
matrix elements from eq.(12);
The curves in this figure are labeled in the same way as in fig.1.

\end{document}